         \documentclass[twocolumn,eqsecnum,aps,prb,showpacs]{revtex4}
         \usepackage[dvips]{graphicx}
\providecommand{\mB}{\mu{\rm B}}
\providecommand{\veck}{\mathbf{k}}

\providecommand{\vecr}{\mathbf{r}}
\providecommand{\veck}{\mathbf{k}}
\providecommand{\FIGONE}{
\begin{figure}
\includegraphics[width=5cm,keepaspectratio]{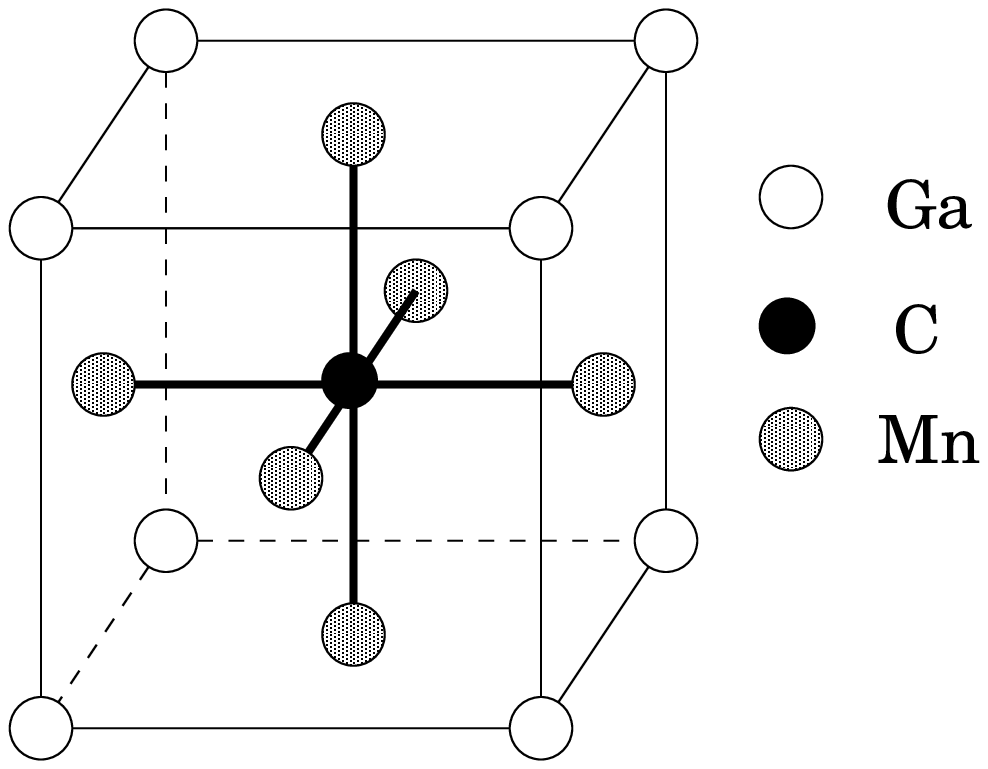}
\caption{Inverse perovskite-type crystal structure.
\label{fig.invperov}
}
\end{figure}
}
\providecommand{\FIGTWO}{
\begin{figure}
\includegraphics[width=6cm,keepaspectratio]{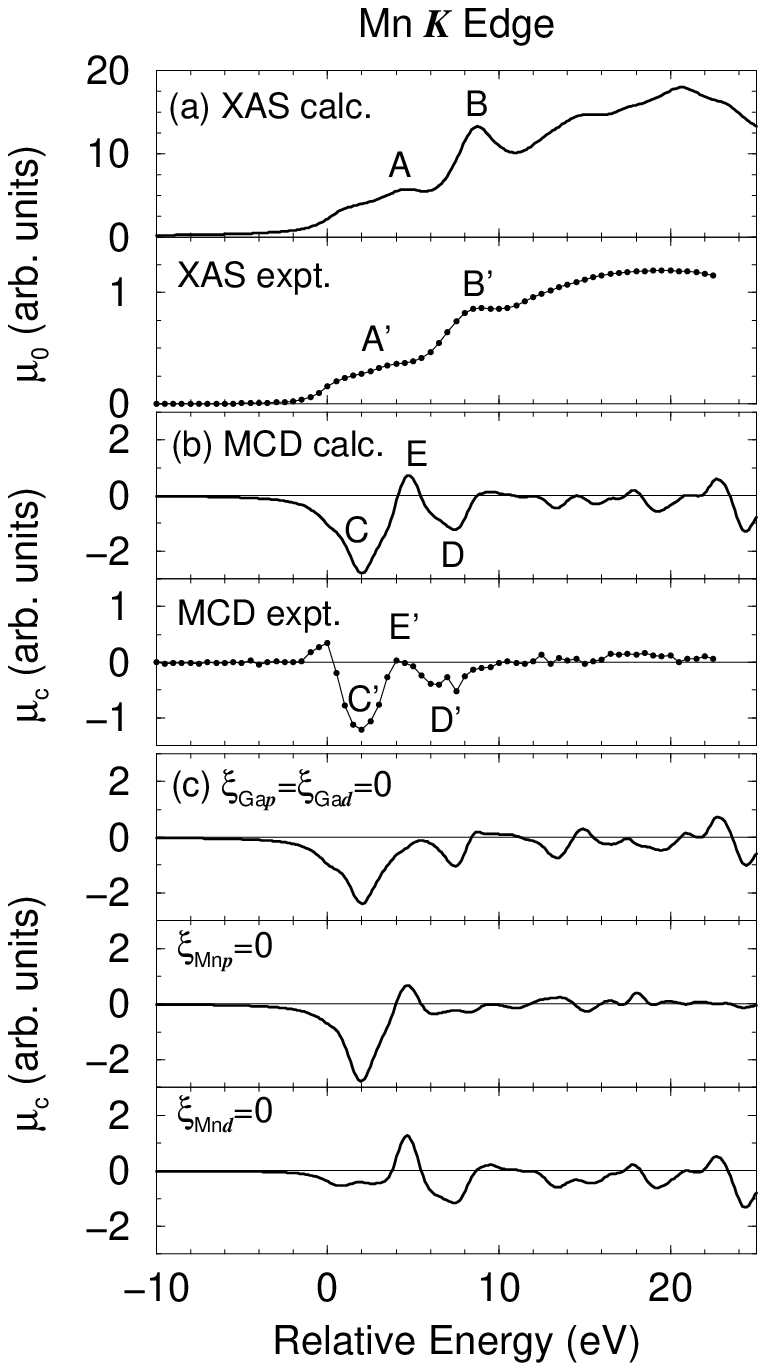}
\caption{
(a)Total absorption coefficient $\mu_0(\omega)$, and (b)XMCD spectra
$\mu_c(\omega)$ at the Mn $K$ edge, as a function of photon energy.
The origin of energy corresponds to the excitation to the Fermi level.
The normalized intensities are magnified to $10^{3}$ times.
The experimental data are for a powdered sample in the ferromagnetic phase
at $T=200$ K\cite{Uemoto}.
(c)XMCD spectra calculated with turning off the SOI on all the states
at Ga sites, on the $p$ symmetric states at Mn sites, and on the $d$ symmetric 
states at Mn sites, respectively. 
$\xi_{{\rm A}b}=0$ means that the SOI on the $b$
symmetric states at A sites is turned off.
\label{fig.Mn}
}
\end{figure}
}
\providecommand{\FIGTHREE}{
\begin{figure}
\includegraphics[width=6cm,keepaspectratio]{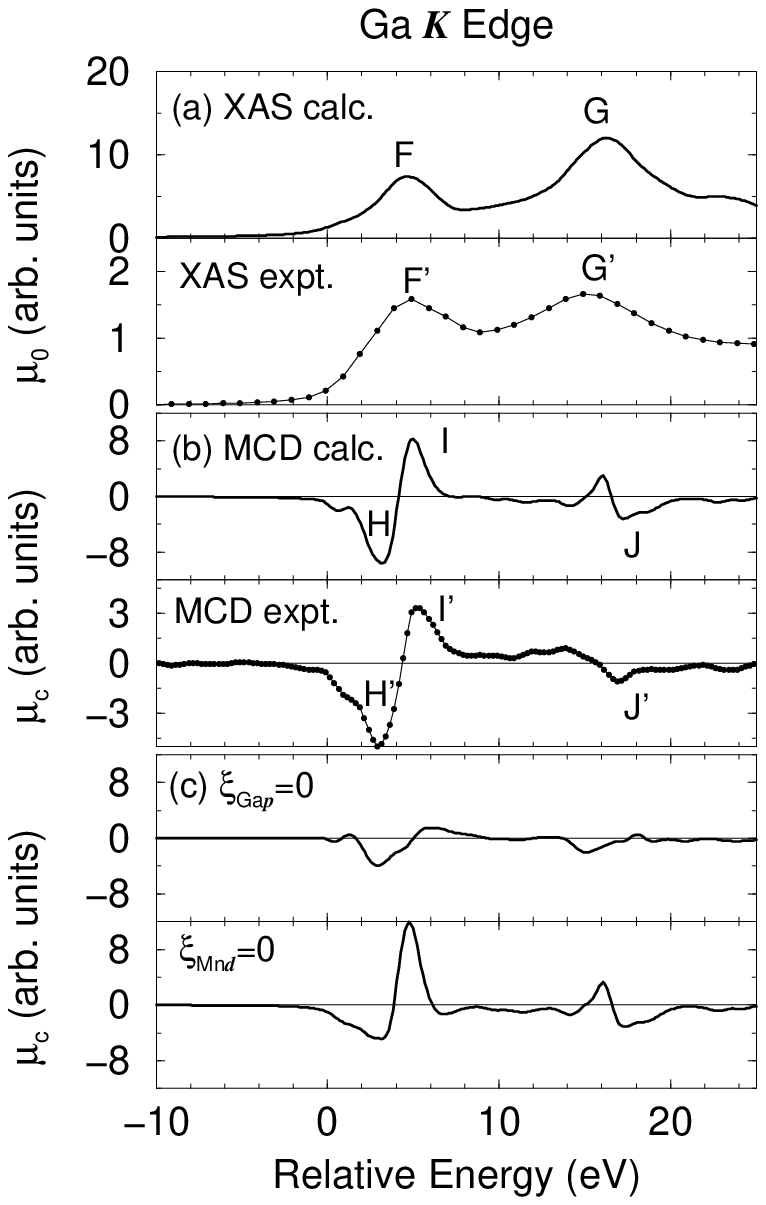}
\caption{
(a)Total absorption coefficient $\mu_0(\omega)$,and (b)XMCD spectra
$\mu_c(\omega)$ at the Ga $K$ edge, as a function of photon energy.
The origin of energy corresponds to the excitation to the Fermi level.
The normalized intensities are magnified to $10^{3}$ times.
The experimental data are for a powdered sample in the ferromagnetic phase
at $T=200$ K\cite{Kawamura}.
(c)XMCD spectra calculated with turning off the SOI on the $p$ symmetric
states at Ga sites, and on the $d$ symmetric states at Mn sites, respectively.
$\xi_{{\rm A}b}=0$ means that the SOI on the $b$
symmetric states at A sites is turned off.
\label{fig.Ga}
}
\end{figure}
}
\providecommand{\FIGFOUR}{
\begin{figure}
\includegraphics[width=6cm,keepaspectratio]{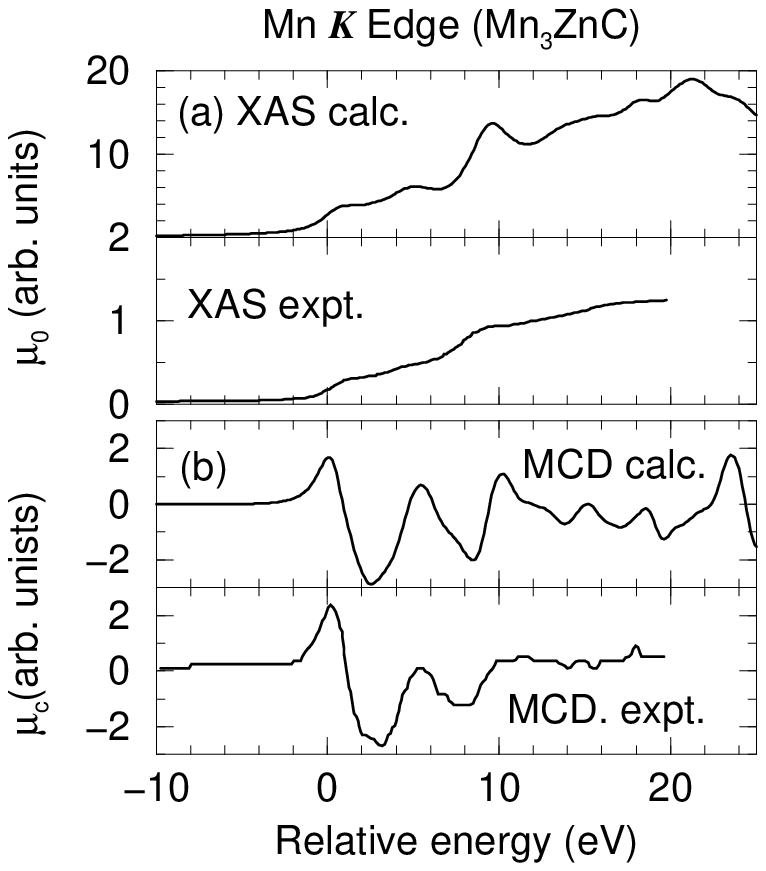}
\caption{
(a)Total absorption coefficient $\mu_0(\omega)$,and (b)XMCD spectra
$\mu_c(\omega)$ at the Mn $K$ edge in Mn$_3$ZnC, as a function of photon energy.
The origin of energy corresponds to the excitation to the Fermi level.
The normalized intensities are magnified to $10^{3}$ times.
The experimental data are for a powdered sample in the ferromagnetic phase
at $T=300$ K\cite{Kawamura}.
\label{fig.Mn_mzc}
}
\end{figure}
}
\providecommand{\FIGFIVE}{
\begin{figure}
\includegraphics[width=5cm,keepaspectratio]{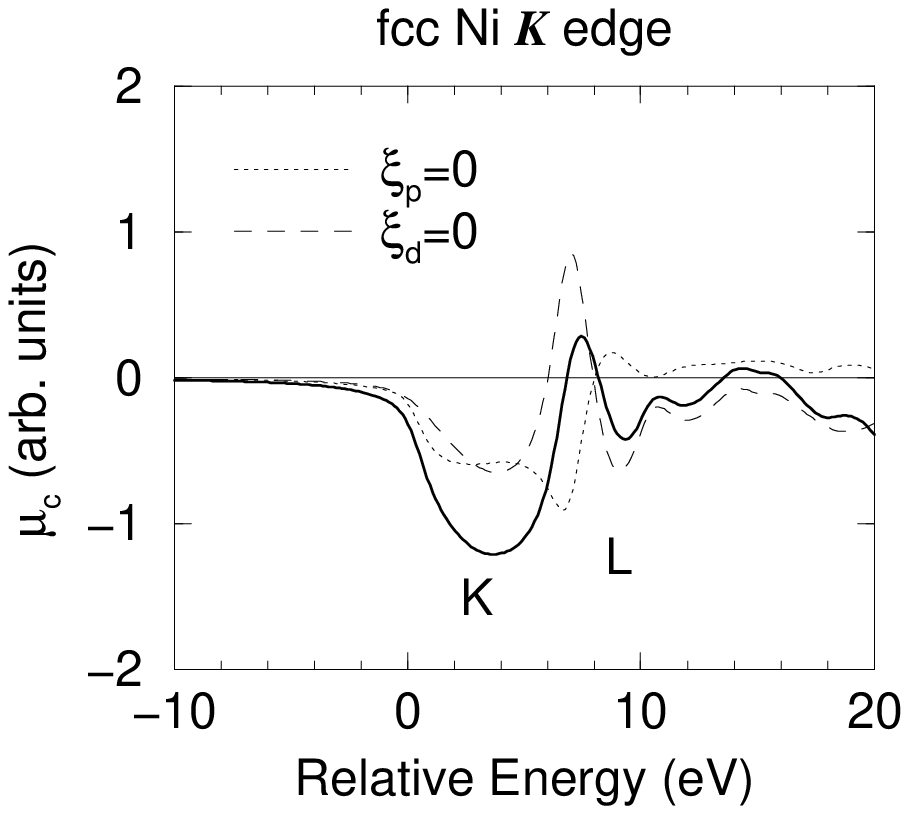}
\caption{
Calculated XMCD spectra at the $K$ edge in the ferromagnetic Ni metal,
as a function of photon energy, which are divided by the total absorption 
coefficient at $\hbar\omega=\epsilon_{\rm F}+20$ eV and magnified to $10^{3}$ times.
The broken and dotted lines are values
calculated with turning off the SOI on the $p$ symmetric states and 
on the $3d$ states, respectively.
\label{fig.Ni}
}
\end{figure}
}
\begin{document}
\title{X-Ray Magnetic Circular Dichroism at the $K$ edge of Mn$_3$GaC}
\author{Manabu Takahashi}
\affiliation{Faculty of Engineering, Gunma University, Kiryu, Gunma 376-8515, Japan}
\author{Jun-ichi Igarashi}
\affiliation{Synchrotron Radiation Research Center,Japan Atomic Energy Research Institute,
Mikazuki, Sayo, Hyogo 679-5148, Japan}
\date{ \today }
%
%--------%---------%---------%---------%---------%---------%---------%---------%
%
\begin{abstract}

We theoretically investigate the origin of the x-ray magnetic circular
 dichroism (XMCD) spectra at the $K$ edges of Mn and Ga in the ferromagnetic
 phase of Mn$_3$GaC on the basis of an ab initio calculation.  Taking
 account of the spin-orbit interaction in the LDA scheme, we obtain the XMCD
 spectra in excellent agreement with the recent experiment.  We have
 analyzed the origin of each structure, and thus elucidated
 the mechanism of inducing the orbital polarization in the $p$ symmetric
 states. We also discuss a simple sum rule connecting the XMCD spectra with 
 the orbital moment in the $p$ symmetric states.

\end{abstract}

\maketitle
%
%--------%---------%---------%---------%---------%---------%---------%---------%
%
%%%%%%%%%%%%%%%%%%%%%%%%%%%%%%%%%%%%%%%%%%%%%%%%%%%%%%%%%%%%%%%%%%%%%%%%%%%%%%%%
%
\section{Introduction}
X-ray magnetic circular dichroism (XMCD) has attracted much interest as a
 useful tool to investigate magnetic states\cite{Schutz1}.  The spectra
 directly reflect the magnetic order at the $L_{2,3}$ edges of
 transition-metal compounds and at the $M_{4,5}$ edges of rare-earth
 compounds, because by photoabsorption the core electron enters the $3d$ or
 $4f$ states which constitute the magnetic order.  Sum rules have been found
 useful to evaluate the orbital magnetization\cite{Chen,Thole,Carra}.
On the other hand, at earlier time, it was not clear where the XMCD spectra
 at the $K$ edge of transition-metal compounds come from, because the $4p$
 states the core electron enters by photoabsorption are not the states
 constituting the magnetic order\cite{Schutz2,Ebert,Stahler}. One of the
 present authors and Hirai have analyzed the XMCD at the $K$ edge 
 of ferromagnetic metals, Fe, Co, and Ni, and have found that the spectra 
 come from the $4p$
 orbital polarization induced by the mixing to {\em the $3d$ states at
 neighboring sites}\cite{Igarashi1a,Igarashi1b}. 
 Subsequent experiments and calculations reached the similar conclusion
 for several transition-metal compounds. \cite{Guo,Ankudinov,Rueff,Brouder,Chaboy96,Chaboy98,Chaboy98b,Yamamoto}

The above mechanism is a consequence of an
 extended character of $4p$ states, and is closely related to the mechanism
 of the resonant x-ray scattering (RXS) at the $K$ edges of transition
 metals in several compounds\cite{Murakami,Noguchi}. The RXS intensity on
 superlattice spots was considered to be a direct reflection of the orbital
 order in LaMnO$_3$\cite{Ishiharaa}, but subsequent studies based
 on band structure calculations have revealed that the RXS intensity arises
 mainly from the lattice distortion which modulates
the $4p$ states in the intermediate state of the second order process of the
 RXS\cite{Elfimov,Benfatto,Takahashi1a}.
 Recently, ab initio calculations
 of the RXS spectra have been carried out for YTiO$_3$ and YVO$_3$, having
 clarified the important role of lattice
 distortion\cite{Takahashi2a,Takahashi2}.

Despite such progress, the underlying mechanism of inducing $4p$ orbital
 polarizations is not fully understood.  The XMCD experimental data recently
 obtained in the ferromagnetic phase of Mn$_3$GaC\cite{Uemoto,Kawamura}
 seems suitable to make clear the mechanism, because considerable signals
 have been found not only at the $K$ edge of Mn but also at the $K$ edge of
 Ga.  In this paper, we analyze the XMCD spectra at the $K$ edges in
 Mn$_3$GaC through an ab initio calculation.  We use the
 Korringa-Kohn-Rostoker (KKR) method within the muffin-tin (MT)
 approximation \footnote{ We add an empty sphere between neighboring Ga
 sites in order to reduce the interstitial volume. The MT radii used in the
 calculation are 0.25 in units of lattice constant so that the MT spheres
 are touching with one another.} in the local density approximation (LDA)
 scheme, and take account of the spin-orbit interaction (SOI).  We neglect
 the core-hole potential in the final state.  We expect that the final-state
 interaction is unimportant except for the Fermi-edge singularity, because
 it is known from the study of the XMCD in Fe, Co, Ni, that the spectra are
 well reproduced without taking account of the final-state interaction
 \cite{Ebert,Stahler,Igarashi1a,Igarashi1b}.  This may come from the fact
 that the $4p$ states have small amplitude inside the MT sphere and thus are
 little subject to the core hole potential.  

In Sec.~II, we calculate the electronic structure in a ferromagnetic phase
 of Mn$_3$GaC. In Sec.~III, we present the formulas for the XMCD spectra
 and derive a sum rule. We also discuss the calculated spectra in comparison
 with the  experiment. Section IV is devoted to concluding remarks.

\section{Electronic structure}
This material takes an ``inverse'' perovskite structure\cite{Fruchart},
 as schematically shown in Fig.~\ref{fig.invperov}. It
 has an interesting magnetic property; the antiferromagnetic phase in low
 temperatures turns into a ferromagnetic phase in high temperatures through
 a first-order transition at $168$ K\cite{Uemoto}. We carry out the band
 calculation assuming a ferromagnetic phase with the magnetization fixed to
 the opposite of the $[111]$ direction, although an antiferromagnetic phase
 may be stabler than the ferromagnetic phase. By fixing the magnetization
 direction, we neglect the small magnetic anisotropy.
We define the density of states (DOS) inside the MT sphere at a Mn site and
 at a Ga site. On Mn sites, the DOS projected onto the $d$ symmetric states
 is dominating near the Fermi level.  The $s$ and $p$ symmetric DOS's are
 found extremely small. The spin and orbital angular momenta, denoted as $S$
 and $L$ in units of $\hbar$, are defined also inside the MT sphere. We get
 $S=0.73$ and $L=0.042$ in the $d$ symmetric states, both of which are
 pointing to the $[111]$ direction, while we get $S=-0.0006$, $L=0.00006$ in
 the $p$ symmetric states (the $-$ sign means that the angular momentum is
 pointing to the opposite of the $[111]$ direction).  The total magnetic
 moment becomes $-(2S+L)\mu_B=-1.50\mu_{B}$ per Mn.  This value is
 consistent with the magnetization measurement\cite{Fruchart} and the
 previous band calculations without taking account of the
 SOI\cite{Shirai,Ivanovskii}.  On the other hand, on Ga sites, the $3d$
 states are located about $12$ eV below the Fermi level. The $s$ and $p$
 symmetric DOS's are found small above the Fermi level.  Small angular
 momenta are induced due to the influence of the magnetic moments at Mn
 sites. We get $S=-0.011$, $L=0.00046$ in the $p$ symmetric states, and much
 smaller values, $S=0.0021$, $L=-0.00041$, in the $d$ symmetric states.
\FIGONE

\section{XMCD spectra}

We calculate the absorption coefficient, neglecting the core hole potential.
 Assuming that photons are propagating along the $[111]$ direction, we have
 the expressions for the right-handed $(+)$ and left-handed $(-)$ circular
 polarizations,
\begin{eqnarray}
 \mu_{\pm}(\omega) &\propto& \sum_{n,\veck}\left|\int r^2 dr\int {\rm d}\Omega 
       \phi_{n,\veck}(\vecr)^* rY_{1,\pm 1}(\Omega)R_{1s}(r )       \right|^2 \nonumber\\
 &\times&\delta(\hbar\omega-\epsilon_{n,\veck})\theta(\epsilon_{n,\veck}-\epsilon_F),
\end{eqnarray}
where the spherical harmonic function $Y_{1,\pm 1}(\Omega)$ is defined with
 the quantization axis along the $[111]$ direction.  The step function
 $\theta(x)$ ensures that the sum is taken over states above the Fermi
 level.  The $R_{1s}$ represents the $1s$ wave function of Mn or Ga, and
 $\phi_{n,\veck}$ represents the wave function with the band index $n$,
 wave-vector $\veck$ and energy $\epsilon_{n,\veck}$.  In the actual
 calculation, we replace $\delta(\hbar\omega-\epsilon_{n,\veck})$ by a
 Lorentzian form $(\Gamma/\pi)/((\hbar\omega-\epsilon_{n,\veck}
 )^2+\Gamma^2)$ with $\Gamma=1$ eV, in order to take account of the $1s$
 core hole life-time width.  The total absorption coefficient and the XMCD
 spectra are defined by 
\begin{eqnarray}
 \mu_0(\omega)&=&[\mu_+(\omega)+\mu_-(\omega)]/2 ,\\
 \mu_c(\omega)&=&[\mu_+(\omega)-\mu_-(\omega)].
\label{eq.xmcd}
\end{eqnarray}

The XMCD intensity arises from the $p$ orbital polarization in the
 unoccupied states, as is obvious from Eq.(\ref{eq.xmcd}).  This quantity
 can be connected with the orbital moment in the $p$ symmetric states as
 follows.  Let the wave function be expanded in the MT sphere as
\begin{equation}
 \phi_{n,\veck}(\vecr) = \sum_{\ell,m,\sigma} f^{n,\veck}_{\ell,m,\sigma}
              \psi^{n,\veck}_{\ell,m,\sigma}(\vecr),
\label{eq.expand}
\end{equation}
where $\psi^{n,\veck}_{\ell,m,\sigma}(\vecr)$ represents the wave
 function projected onto the state with the angular momentum $\{\ell,m\}$,
 and spin $\sigma$, and normalized inside the MT sphere.  Then the orbital
 angular momentum in the $p$ symmetric state is expressed as 
\begin{equation}
 L_{\rm p} = \sum_{n,\veck,\sigma}
  (|f^{n,\veck}_{1,1,\sigma}|^2 - |f^{n,\veck}_{1,-1,\sigma}|^2)
  \theta(\epsilon_{\rm F}-\epsilon_{n,\veck}),
\end{equation} 
where the sum is taken over the occupied states. Note that, if the sum is
 taken over all the states both occupied and unoccupied, one may get
 $\sum_{n,\veck,\sigma} (|f^{n,\veck}_{1,1,\sigma}|^2 - |f^{n,\veck}
_{1,-1,\sigma}|^2)=0$.  On the other hand, the absorption coefficient
 may be rewritten as
\begin{eqnarray}
 &&\mu_{\pm}(\omega)\propto\sum_{n,\veck,\sigma}
 \delta(\hbar\omega-\epsilon_{n,\veck})\theta(\epsilon_{n,\veck}-\epsilon_{\rm F})
 |f^{n,\veck}_{1,\pm 1,\sigma}|^2 \nonumber \\
 &&\times \left|\int r^2 dr\int {\rm d}\Omega \psi^{n,\veck}_{1,\pm 1,\sigma}(\vecr)^* 
    rY_{1,\pm 1}(\Omega)R_{1s}(r )\right|^2.
 \label{eq.abex}
\end{eqnarray}
The last factor represents the transition probability from the $1s$ state
 to the $p$ symmetric states, which varies little with respect to the energy
 of the $p$ symmetric states above the Fermi level up to $\sim 20$ eV
 (within $\sim 10\%$ variation). This is because the $p$ symmetric states
 are well approximated by the atomic $4p$ wave function on the region close
 to the center of atoms.  Replacing the last factor by the value at the
 Fermi energy (denoted as $|S_{\rm F}|^2$) in Eq.(\ref{eq.abex}), we obtain
 approximately the relation,
\begin{equation}
 \int^{\omega_2} \mu_c(\omega){\rm d}\omega \propto -|S_{\rm F}|^2 L_{\rm p}.
\label{eq.sum}
\end{equation}
We need to introduce some upper cutoff $\omega_2$, because the transition
 probability is replaced by a constant $|S_{\rm F}|^2$.  It may be set by
 $\hbar\omega_2=\epsilon_{\rm F}+20$ eV.  This sum rule is an extension of
 the relation previously derived on the basis of the tight-binding model for
 Fe, Co, Ni metals\cite{Igarashi1a,Igarashi1b}.

Now we discuss the calculated spectra at the Mn $K$ edge.  Figure
 \ref{fig.Mn}(a) and (b) show the total absorption coefficient 
 $\mu_0(\omega)$ and the XMCD spectra $\mu_c(\omega)$ in comparison with the
 experiment\cite{Uemoto}. The experimental data are for a powdered sample
 under a small magnetic field $H=0.6$ T at $200$ K. The sample is in the
 ferromagnetic phase with the magnetization $\sim 0.8\mB$ per Mn, which
 value is smaller than the calculated one $1.5\mB$.  The difference may be
 ascribed to a temperature effect, which usually reduces the magnetization.
  The calculated value of $\mu_c(\omega)$ is divided by the value of
 $\mu_0(\omega)$ at $\hbar\omega=\epsilon_{\rm F}+20$ eV, while the
 experimental XMCD spectra are divided by the value of the total absorption
 coefficient at the energy about $40$ eV higher than the threshold.  Both
 $\mu_0(\omega)$ and $\mu_c(\omega)$ are in good agreement with the
 experiment.  For the total absorption coefficient, peaks A and B
 correspond well with shoulders A' and B' in the experimental curve.
  For the XMCD spectra, dips C, D, and peak E correspond well with the
 experimental ones C', D', and E'.  Only one deviation from the
 experiment is that the small peak at the Fermi level is not reproduced.
  The intensity integrated up to $\omega_2$ is found negative.  This is
 consistent with the sum rule, Eq.(\ref{eq.sum}), because the value of $L_{\rm
 p}$ is positive at Mn sites. 
 \FIGTWO

The XMCD spectra come from the orbital polarization in the $p$ symmetric
 states, which may be induced by (i) the spin polarization in the $p$
 symmetric states through the SOI, and (ii) the orbital polarization at
 neighboring sites through hybridization.  For making clear the inducing
 mechanism, we calculate the XMCD spectra at Mn sites with turning off the
 SOI on several specified states. The top panel among three panels of
 Fig.~\ref{fig.Mn}(c) shows the XMCD spectra with turning off the SOI on all
 the states at Ga sites.  The spectra remain similar except for a slight
 suppression of peak E, indicating that the orbital polarization at Ga sites
 have little influence on the XMCD spectra at Mn sites.  The middle panel in
 Fig.~\ref{fig.Mn}(c) shows the XMCD spectra with turning off the SOI only
 on the $p$ symmetric states at Mn sites.  Dip C keeps the similar shape,
 while dip D almost vanishes.  This indicates that the $p$ orbital
 polarization corresponding to dip D is induced by the spin polarization in
 the $p$ symmetric states through the SOI.  The bottom panel shows the XMCD
 spectra with turning off the SOI only on the $d$ symmetric states at Mn
 sites.  The intensity of dip C is drastically reduced, but dip $D$ and peak
 E remains similar, indicating that the $3d$ orbital polarization gives rise
 to dip C.  Within the MT approximation, the $3d$ orbital polarization
 cannot polarize the $p$ orbital in the same Mn site, because the $p$-$d$
 Coulomb interaction is spherically averaged inside the MT sphere. Thus we
 conclude that the $3d$ orbital polarization at {\em neighboring} Mn sites
 induces the $p$ orbital polarization corresponding to dip C through the
 $p$-$d$ hybridization.

Next we discuss the XMCD at the Ga $K$ edge.  Figure \ref{fig.Ga}(a) and (b)
 shows the calculated $\mu_0(\omega)$ and $\mu_c(\omega)$ 
 in comparison with the
 experiment\cite{Kawamura}. Both $\mu_0(\omega)$ and $\mu_c(\omega)$ are in
 good agreement with the experiment. For the total absorption coefficient,
 peaks F and G correspond well with the experimental ones. Also, for the
 XMCD spectra, dip H, peak I and dip J correspond well with the
 experimental ones.  The intensity integrated up to $\omega_2$ is found
 negative.  This is again consistent with the sum rule, because the value of
 $L_{\rm p}$ is positive at Ga sites.
 \FIGTHREE
 
 For making clear the mechanism, we
 also calculate the XMCD spectra at Ga sites with turning off the SOI on
 several specified states.  The top panel in Fig.~\ref{fig.Ga}(c) 
 shows the XMCD spectra with turning off the SOI 
 on the $p$ symmetric states at Ga
 sites.  Peak I and dip J are almost suppressed, indicating that the $p$
 orbital polarization corresponding to those structures is induced by the
 spin polarization in the $p$ symmetric states through the SOI. On the other
 hand, dip H is half suppressed, indicating that the above mechanism works
 in a small part on dip H.  The bottom panel shows the XMCD spectra with
 turning off the SOI on the $d$ symmetric states at Mn sites. Dip J
 remains similar, dip H is half suppressed, and peak I is even enhanced.
  This indicates that the negative values are added to both dip H and peak
 I due to the coupling to the $3d$ orbital polarization at neighboring Mn sites.

As shown above, the calculation reproduces well the experimental curves.
 Only the exception is the small positive peak $W$ at the Fermi level on Mn 
 sites in the XMCD spectra (see Fig.~2). We might take it a small error, 
 since the peak is small. However, in case of Mn$_3$ZnC,
 the similar positive peak at the Fermi level becomes large on Mn sites.
 To gain confidence in our calculation, we carry out a similar calculation 
 for Mn$_3$ZnC, assuming a ferromagnetic phase. We get $S=1.088$,
 $L=0.019$ in the $d$ symmetric states, and  $S=-0.0003$,
 $L=-0.00005$ in the $p$ symmetric states on Mn site.
 Figure \ref{fig.Mn_mzc} shows the calculated absorption and XMCD spectra at Mn $K$
 edge, in comparison with the experiment at $T=300$ K
 (ferromagnetic phase). The positive peak at the Fermi level is 
 clearly reproduced.
\FIGFOUR

\section{Concluding remarks}
We have reported an ab initio calculation of the XMCD spectra at
 the $K$ edges of Mn and Ga in the ferromagnetic phase of Mn$_3$GaC, by
 taking account of the SOI in the LDA scheme.  The calculated spectra show
 excellent agreement with the recent experiment.  The spectra have
 explicitly been shown to arise from the $p$ orbital polarization, and the
 mechanism of its induction has been fully analyzed; the associated process
 is identified for each structure.  The present result may serve a guide to
 analyze the $K$ edge XMCD spectra in other materials.  Such usefulness has
 been demonstrated for the ferromagnetic Ni metal.  We have also derived a
 simple sum rule, which connects the XMCD spectra with the orbital moment, and
 have confirmed it works. Judging from the agreement of the calculated 
 spectra with the experiment, we think that the effect of the 1s-core-hole 
 potential in the final state is negligibly small.

The present finding is rather general and can be applied to other systems.
 To demonstrate this, we compare the XMCD spectra in Mn$_3$GaC with those 
 in the ferromagnetic Ni metal. Figure \ref{fig.Ni} shows the latter quantity
 calculated with the same method as above. We divide $\mu_c(\omega)$ by
 $\mu_0(\omega_2)$ at $\hbar\omega_2=\epsilon_{\rm F}+20$ eV.  A large dip
 K appears above the Fermi level; it is half suppressed if the SOI is
 turned off either in the $p$ symmetric states or in the $3d$ states.  This
 behavior is similar to dip C in Mn$_3$GaC, although the contribution of
 the SOI in the $p$ symmetric states is a little larger in Ni.  On the other
 hand, a tiny dip L at high energy disappears without the SOI in the $p$
 symmetric states, similar to dip D in Mn$_3$GaC. 
 \FIGFIVE

Finally we point out that this type of ab initio calculation can be applied
 to analyzing the {\em magnetic} RXS spectra at the $K$ edge in
 transition-metal compounds\cite{Neubeck,Igarashi2}, and may clarify 
 the $p$ orbital polarization in {\em antiferromagnetic} phases.
 According to our recent calculation of the magnetic RXS spectra in NiO\cite{usuda},
 the main peak intensity almost disappears when the SOI on the
 4p band is turned off. This finding is consistent with the above analysis,
 because the $4p$ states are located at $\sim 10$ eV higher than 
 the Fermi level.

\begin{acknowledgments}
We would like to thank N. Kawamura and H. Maruyama for providing us the MCD
 experimental data prior to publication and for valuable discussion.  This
 work was partially supported by a Grant-in-Aid for Scientific Research from
 the Ministry of Education, Science, Sports and Culture, Japan.
\end{acknowledgments}
%
%--------%---------%---------%---------%---------%---------%---------%---------%
%
                             \bibliography{p3b}
%
%--------%---------%---------%---------%---------%---------%---------%---------%
%
\end{document}